\documentclass[useAMS,usenatbib]{mnras}
\usepackage{epsfig,appendix,array,rotating,pdflscape,array,longtable,colortbl,appendix}
\usepackage{multirow,url}
\usepackage{booktabs}
\usepackage{lscape, amsmath} 
\usepackage{paralist,subfigure}

\usepackage{relsize}
\usepackage{fixltx2e}
\usepackage{amssymb}
\usepackage{xtab}

\newcommand{\hi}{H\,{\sc i}}

\newcommand{\muJy}{${\mu}$Jy}

\newcommand{\km}{km\,s$^{-1}$}
\newcommand{\degree}{$^{\circ}$}
\newcommand{\halpha}{H${\alpha}$}
\newcommand{\han}{H${\alpha}$ + [NII]}
\newcommand{\msolar}{M$_{\odot}$}

\newcommand{\mhi}{M$({\mathrm {HI}})$}

\newcommand{\angs}{${\mathrm {\AA}}$}
\newcommand{\msolaryr}{M$_{\odot}$\,yr$^{-1}$}
\newcommand{\sig}{$\sigma$}
\newcommand{\ergs}{erg\,cm$^{-2}$\,s$^{-1}$}
\newcommand{\oab}{12+log(O/H)}
\newcommand{\GG}[1]{}
\title{Arp\,202: a TDG formed in a parent's extended dark matter halo?}
\author[Scott {\it{et al.}}]{T. C. Scott$^{1}$, P. Lagos$^{1}$, S. Ramya$^{2}$, C. Sengupta $^{3}$, S. Paudel$^{4}$,  D. K. Sahu$^{5}$, \and K. Misra$^{6}$, 
J. -H. Woo $^{4}$,  and B. W. Sohn$^{4}$   \\ 
$^{1}$ Institute of Astrophysics and Space Sciences (IA), Rua das Estrelas, 4150-762 Porto, Portugal\\
$^{2}$ Shanghai Astronomical Observatory, 80 Nandan Road, Shanghai 200030, China \\
$^{3}$ Department of Astronomy, Yonsei University, 50 Yonsei-ro, Seodaemun-gu, Seoul, Republic of Korea\\
$^{4}$ Korea Astronomy and Space Science Institute, 776, Daedeokdae ro, Yuseong gu, Daejeon, 305-348, Republic of Korea\\ 
$^{5}$ Indian Institute of Astrophysics, Koramanagala, Bangalore 34, India\\
$^{6}$ Aryabhatta Research Institute of Observational Sciences (ARIES), Manora Peak, Nainital-263002, Uttarakhand, India\\
}

\begin{document}

\date{Received  ; accepted  }
\date{}
\pagerange{\pageref{firstpage}--\pageref{lastpage}} \pubyear{}

\maketitle

\label{firstpage}

\begin{abstract}
 We report on \han\  imaging of the Arp\,202 interacting pair and its tidal dwarf galaxy (TDG) candidate as well as a GMOS long slit spectrum from the TDG candidate, observed  with the   Gemini North telescope. Our \han\ imaging reveals the TDG to have an elongated  structure,  $\sim$ 1.9 kpc in length with the two principal star forming knots at either end.   Our observations also show the TDG candidate has  a recessional V$_{H\alpha}$ $\sim$ {3032 \km, within 100 \km\ of the parent pair's mean velocity and  an oxygen abundance  of 12+log(O/H) = 8.10$\pm$0.41. The TDG's oxygen abundance is in good agreement with that of a  star forming  region in NGC\,2719A, one of the parent galaxies,  which has an estimated oxygen abundance of 12+log(O/H)  = 8.05$\pm$0.41. The TDG's V$_{H\alpha}$ and oxygen abundance confirm previous results validating the candidate as a TDG. The absence of detectable emission from the TDG  in  \textit{Spitzer}  3.6 $\mu$m and 4.5 $\mu$m images together with the lack of absorption lines and weak continuum in the spectrum is consistent with absence of an old population  ($\gtrsim$ 0.5 Gyr). The location of the TDG within the  interaction debris and the  absence of indicators of an old  stellar population in the TDG are  consistent with a scenario in which  the TDG is formed from \hi\ stripped from the parent galaxies and within the  extended dark matter  halo of one of the parents as proposed by \citep{bournaud03,duc04}. }

\end{abstract}

\begin{keywords}
galaxies: spiral - galaxies: interactions - galaxies: kinematics and dynamics - 
galaxies: individual: Arp 202 \halpha: galaxies - spectra: galaxies \textcolor{black}{galaxies: tidal dwarfs}
\end{keywords}
\section{Introduction}
\label{intro}

Pre-merger tidal interactions between pairs of approximately equal mass  galaxies, where at least one of them is gas rich, can result in large masses of stars and gas  being  tidally stripping from the parent galaxies'  gas  and stellar disks \citep[e.g.][]{struck99,hibbard05,duc99,seng13}. Most of this tidally stripped material  will eventually fall back into the potential of one or other of the pair, or at later stages the new merged galaxy.  However, during  this  process  self-gravitating bodies,  with a range of masses, are observed to  form within the tidal debris  \citep[e.g.][]{hancock09,smith10a}. At the top  end of their mass range are bodies, with a gas content and luminous mass similar  to dwarf galaxies ($\sim$ 10$^9$\msolar) known as tidal dwarf galaxies (TDGs) \citep{duc99,duc00,braine01,duc12}. TDG formation models propose the collapse of the \hi\  debris into  \hi\ disk, parts of which  convert to the molecular phase and further collapse to generate  in--situ star formation \citep{bournaud08}.

TDG formation models \citep[e.g.][]{dabring13} also  predict TDGs will  contain little or no dark matter \textcolor{black}{(DM)}, although this has proven to be difficult to confirm observationally both because their stellar and gas  \textcolor{black}{masses} are subject to large uncertainties \cite[][]{lelli15}  and  it is not clear whether the assumption that these components are  virialised, on which calculations of their M$_{dyn}$ rely, is valid.  \cite{flores16} recently pointed out many observed TDGs  are probably too young  for their gas  and  star forming clumps to have virialised. Modelling, \textcolor{black}{assuming the absence of a  \textcolor{black}{ TDG} DM} halo,  also shows that  supernovae and stellar wind feedback, following an initial star formation (SF) burst, is insufficient to break up a TDG  \citep{ploeck14}.

Because TDGs are predicted to form from chemically enriched gas originating in  the parent disk(s)  \textcolor{black}{TDG} metallicities, indicated by oxygen abundance, are expected to be approximately solar \citep[\oab\ = 8.69$\pm$0.05,][]{asplund2009} rather than  the significantly sub--solar \textcolor{black}{metallicities}  observed in normal dwarf galaxies \citep[\oab\ = 7.4 to 7.9,][]{lee03}. As a result metallicity is one of the principal discriminants used to distinguish  TDGs from normal \textcolor{black}{dwarf galaxies}.

Modelling by \cite{bournaud03} and \cite{duc04} indicates that for a parent galaxy with a truncated DM halo\textcolor{black}{, self gravitating bodies will  form in the interaction debris  and be  distributed along the length of the tidal tail. However  masses of these bodies will be lower than a TDG.} In contrast, for a  parent galaxy with an extended DM halo,  a large gas mass can be  carried away from the parent and  accumulate at the  extremity of the tidal tail. This accumulated gas at the end of tidal tail can subsequently gravitationally collapse to form a TDG.  Cases where a TDG forms from  pure gas debris collapse, as opposed to gas collapse \textcolor{black}{ in conjuction with }  a significant mass of stellar debris, were proposed as a separate class of TDG in \cite{duc04}. TDG formation under the extended DM halo \textcolor{black}{--} pure gas collapse scenario implies both  in--situ star formation  and stellar  population \textcolor{black}{ages  exclusively  younger  than the age of the}  parent interaction that gave rise to the TDG. The \hi\ debris distribution and TDG location in Arp\,105 and Arp\,181 appear consistent with this scenario \citep{duc97,seng13}. \textcolor{black}{In the case of the Arp\,105 TDG a study by  \citep{Boquien10} concluded it appeared ``devoid of stellar populations older than 10$^9$ years". } The TDGs surrounding NGC\,5291 \citep{duc98,fensch16} are likely \textcolor{black}{to be} examples of the pure gas collapse type\textcolor{black}{,  within a large scale \hi\ debris ring formed following the collision of two galaxies}. 

\halpha\ and ultraviolet (UV)  emission, tracing young SF on time scales of 10$^7$\,yr and 10$^8$\,yr respectively \citep{bosel09}, have been observed within evolving tidal debris, including in TDG candidates and lower mass self gravitating SF clumps \citep[e.g.][]{neff05,hancock09,smith10a}.  As a tracer of the most  recent  \textcolor{black}{SF}, \halpha\ is well suited, from a time scale view point,  for detecting the in--situ \textcolor{black}{SF}  predicted for TDGs. \textcolor{black}{Studies of } \halpha\  and UV emission  can therefore assist with the validation of \textcolor{black}{TDGs.} 

\begin{figure*}
\begin{center}
\includegraphics[scale=0.7]{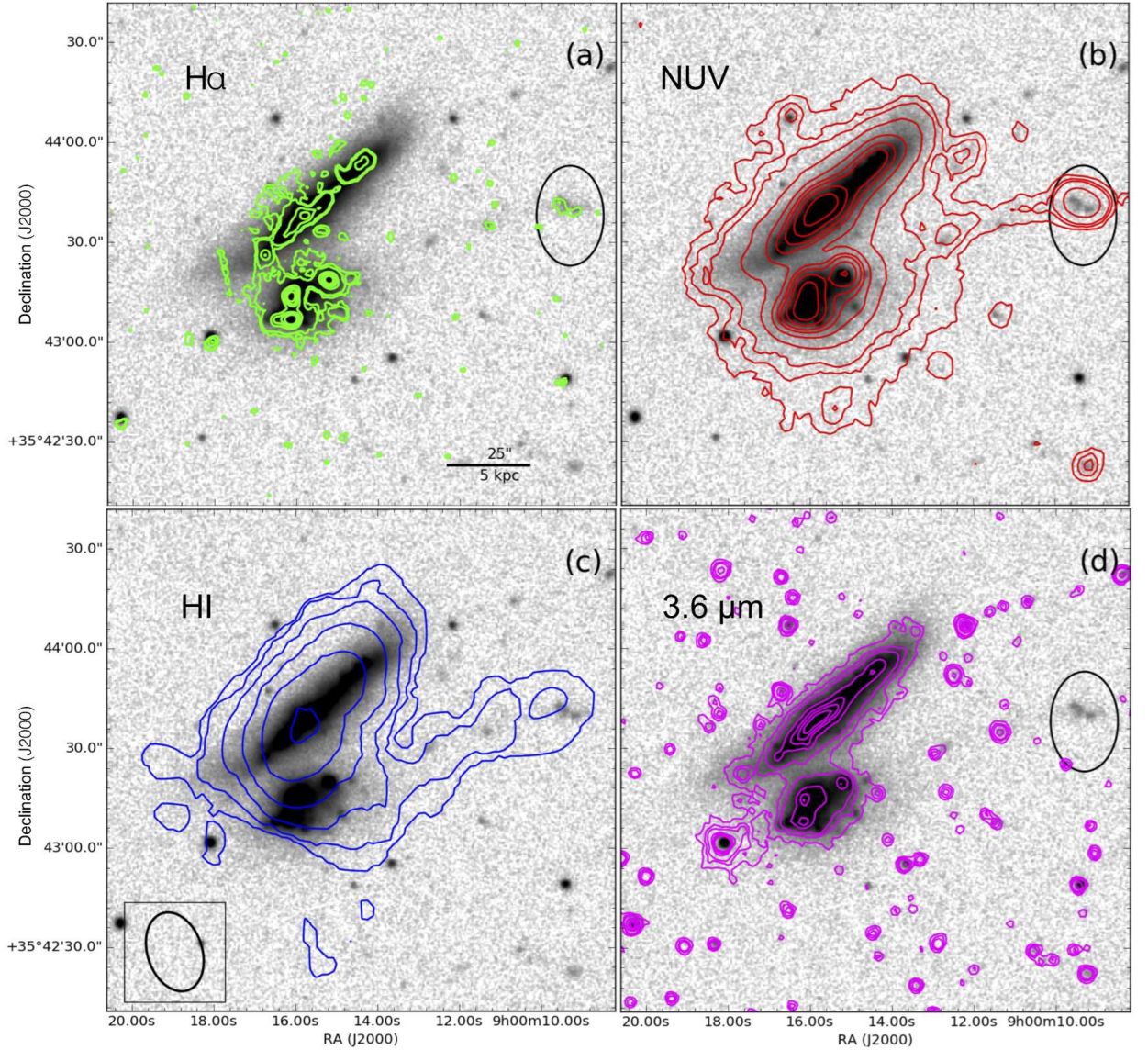}
\caption{\textbf{Arp\,202:}  \textit{\textbf{(a):}} \han\ contours from a smoothed  (boxcar 6 $\times$ 6)  HCT image. The first contour is at 3 $\sigma$ above the background level in the smoothed image. The  resolution of the contour image is $\sim$ \textcolor{black}{1.7} arcsec.  \textit{\textbf{(b): }}  Contours  from an  NUV (\textit{GALEX}) image, with the first contour at 3 $\sigma$ above the background level. The  resolution of the contour image is $\sim$ \textcolor{black}{4} arcsec. \textit{\textbf{(c):}} \hi\  column density contour levels at  10$^{20}$ atoms /cm$^2$ $\times$ (4.4, 7.5, 14.5, 26.0, 43.4, 66.6). The ellipse within the box indicate the size and orientation of the GMRT \textcolor{black}{full width half power (FWHP)} synthesised beam (23.4 arcsec $\times$ 16.3 arcsec). \textit{\textbf{(d):}} Contours from a \textit{Spitzer} 3.6 $\mu$m image, with the first contour at 3 $\sigma$  above the background level. The  resolution of the contour image is $\sim$ \textcolor{black}{2} arcsec. In each panel the black ellipses indicate the region in which TDG candidate is projected. For all panels the background is  a SDSS g --  band image.}
\label{fig1} 
\end{center}
\end{figure*}

In this paper we present   \han\ imaging from the Himalayan Chandra Telescope\footnote{\url{https://www.iiap.res.in/iao_telescope}} (HCT) of the \textcolor{black}{Arp\,202 interacting galaxy pair}   (NGC\,2719 and NGC\,2719A) and  their TDG candidate (Figure \ref{fig1}), together with a \textcolor{black}{Gemini Multi--Object Spectrograph (GMOS) } optical spectrum  for  the  TDG candidate. Previous  UV (\textit{GALEX}) observations revealed a  very  blue (FUV - g = -- 0.47)   diffuse object at the end \textcolor{black}{of} a tidal tail eminating from NGC\,2719A \citep{smith10a}, which the authors classified as a TDG candidate. However,   \han\ imaging with the SARA\footnote{Southeastern Association for Research in Astronomy}  0.9-m telescope at Kitt Peak \citep{smith10b} and   the \textit{Spitzer} at  8 $\mu$m imaging, reported in  \cite{smith07}, failed to \textcolor{black}{detect}  counterparts for the TDG \textcolor{black}{candidate.}     \cite{smith10b} reported the TDG  was at the same redshift as the Arp\,202 pair with an oxgen abundance   \oab\  $\sim$ 8.9, which is  above the \oab\ mean value of the \cite{duc99} TDG sample (\oab\ $\sim$ 8.5). As far as we are aware  the spectrum from which the \textcolor{black}{Arp\,202 TDG's redshift and}  metallicity was derived remains unpublished. Our  GMRT \hi\ mapping of Arp\, 202 \citep{seng14}  revealed  an \hi\ tidal tail, which emanates from  NGC\,2719A and  extends to  the projected position of the TDG candidate. The \hi\ mapping provided \textcolor{black}{morphological and kinematic}   evidence linking the interacting pair, the  \hi\ tidal tail and the  \hi\  TDG  counterpart. \textcolor{black}{Table  \ref{table_1} summaries the properties of the TDG candidate.} 

\cite{schecht2012} presented a FUV –- g vs g –- r plot (their figure 11) for the UV detected TDG candidates from  \cite{smith10a} and two other candidates, Holmberg IX and NGC\,4656UV.  Arp\,202's  TDG candidate appears in their plot as one of the candidates with extreme blue colour, together with NGC\,4656UV, Arp 305 and Holmberg IX, which is one of the strongest TDG candidates \citep{sabbi08}. In addition to our \hi, the HCT \han\ imaging   and the GMOS spectrum, we utilised Sloan Digital Sky Survey (SDSS), \textit{Spitzer} and Galaxy Evolution Explorer (\textit{GALEX}) public archive data and images to help understand the relation between the gaseous and  stellar components of the Arp\,202 system. 

The paper is arranged as follows: Section \ref{obs} gives details of the HCT and GMOS observations,  with  observational results in section \ref{results}.  A discussion follows in section \ref{dis} with concluding remarks in section \ref{summary}. The average of the \textcolor{black}{optical}  radial velocities of NGC\,2719 and NGC\,2719A is 3097 \km. Using this average velocity and assuming H$_0$ = 75 \km\ Mpc$^{−1}$, we adopt a distance of 41.3 Mpc to NGC\,2719 and NGC\,2719A and the TDG. At this distance the spatial scale is $\sim$12 kpc \textcolor{black}{ arcmin$^{-1}$}. J2000 coordinates are used throughout the paper, including in the figures.  

\begin{table}
\centering
\begin{minipage}{75mm}
\scriptsize
\caption{Arp\,202 TDG candidate properties}
\label{table_1}
\begin{tabular}{llrr}
\hline
Property\footnote{From this paper (bold text) otherwise from \cite{seng14}}&Units&value \\ 
\hline
\hline
V$_{radial(HI)}$&[\km]&3047 $\pm$13.6  \\
\textbf{V$_{radial (H\alpha)}$}&\textbf{[\km]}& \textbf{\textcolor{black}{3032}  }\\

\textbf{RA}&\textbf{[h:m:s]}&\textbf{09:00:09.57}\\
\textbf{DEC}&\textbf{[d:m:s]}&\textbf{+35:43:42.15}\\
Distance&[Mpc]&  41.3 \\
M$_{HI}$ &10$^8$ \msolar &1.0$\pm$0.5\\
\textbf{M$_*$\footnote{\textcolor{black}{Mean of M* calculated from SDSS g -- r, r -- i, and g -- i  band colours from \cite{smith10a} using the formulae from \citep{bell03} and SDSS band solar luminosity  parameters from \cite{blanton03}.} }}&\textbf{10$^8$ \msolar }&\textbf{\textcolor{black}{0.4$\pm$0.2}}\\
M$_{dyn}$&10$^8$ \msolar &3.9\\
\textbf{Size(\halpha) major /minor } &\textbf{[arcsec]}&  \textbf{4.6 x 1.3 } \\
\textbf{Size(\halpha) major /minor}& \textbf{[kpc]}& {\bf \textbf{1.917 x 1.295}} \\
Colour& FUV -- g&-0.47    \\
\textbf{Metallicity}&\textbf{[12+log(O/H)]}&\textbf{\textcolor{black}{8.10\textcolor{black}{$\pm$0.41}}}\\
\textbf{F$_{Ha}$}&\textbf{[erg cm$^{-2}$s$^{-1}$]}&\textbf{\textcolor{black}{2.62}$\pm$\textcolor{black}{.003} $\times$ 10$^{-15}$}\\
\textbf{L$_{Ha}$}&\textbf{[erg s$^{-1}$]}&\textbf{\textcolor{black}{9.45} $\times$ 10$^{38}$}\\
\textbf{SFR(\halpha) }&\textbf{[10$^{-3}$\msolaryr]}&\textbf{\textcolor{black}{7.0$\pm$1}}\\
\hline
\end{tabular}
\end{minipage}
\end{table}

\section{Observations}
\label{obs}

\subsection{\han\ imaging and data reduction}
\label{obs-hct}
\han\  and $R$ -- band observations of the Arp\,202 system were obtained   with the 2m \textcolor{black}{HCT}, Hanle, India. Observations were carried out  on October 9th 2015. Imaging was carried out  using  Himalaya Faint Object Spectrograph Camera (HFOSC), equipped with a $2K\times4K$ SITe CCD chip, but only the central $2K\times2K$ region (field of view of $10\times10$ arcmin$^2$) was used for imaging. The plate scale was 0.296 arcsec\,pixel$^{-1}$.  Multiple short exposures of Arp\,202 were obtained \textcolor{black}{with} \han\  and Bessel $R$ -- band filters which were then median combined to give  total on target  integrations of 55 min and 20 min, respectively. \textcolor{black}{ The \han\ filter was  centred at a  wavelength of  6563 \angs\  with a bandwidth of 100\angs.  At the redshift of Arp\,202  emission from the [NII] 6549.9 \angs\, [NII] 6585.3 \angs\ and \halpha\ 6564.6 \angs\ lines fell within the filter's bandwidth.} The CCD calibration frames, namely, bias frames were observed throughout the night at about $\sim$ 1 hr intervals.  Morning and evening twilight flat fields were observed and used for pixel-to-pixel correction. The \textcolor{black}{ observations were carried  out under spectroscopic conditions, although at times through passing clouds and therefore} the measured \han\ fluxes are lower limits.

The \textcolor{black}{HCT} data was reduced   in the standard way using  IRAF{\footnote{Image  Reduction \& Analysis Facility 
(IRAF) software distributed by National Optical Astronomy Observatories, which are operated by the Association of
Universities for Research in Astronomy, Inc., under co-operative agreement with the National Science Foundation}} software package tasks. The reduction  included bias-subtraction, flat-field correction and 
alignment of the frames using IRAF tasks {\sc geomap} and {\sc geotran}. The point spread function (PSF) of \han\   image was matched to that of the $R$ -- band image using the IRAF {\sc gauss} task. The sky background in  individual target frames was estimated, from regions away from the galaxies and not affected by stars, which was then subtracted from the target frames. Flux calibration was carried out using the night's observations  of  the spectrophotometric standard star Feige 110 \citep{oke1990}.
The \han\  line image was obtained after subtracting the PSF matched and scaled version of the \textit{R} -- band 
image (which includes some line emission) from the \han\ image as described by \cite{waller1990} and  similar 
to the procedure adopted in \cite{ramya2007,ramya2009}. \textcolor{black}{ The  flux conversion factor  was estimated at 9.077 $\times$ 10$^{-16}$ erg cm$^2$s$^{-1}$ / (count sec$^{-1}$) which is similar to the value obtained in \cite{ramya2007,ramya2009}.} As noted by \cite{james2004}  scaled \textit{R} - band exposures  give  excellent continuum subtraction for observations made during dark nights. \textcolor{black}{The  FWHM of the point spread function, estimated from the  stars in the R band image, is $\sim$ 5.6 pixels (1.7 arcsec).}



\subsection{GMOS spectrum of the TDG}
\label{obs-gem}
A long slit spectrum for the  Arp\,202 TDG candidate  was obtained using the  Gemini Multi--Object Spectrograph (GMOS) on the  8.1 m Gemini North telescope  between  25 November 2015 and 8 January 2016 (Project GN-2015B-Q-65). The on target integration \textcolor{black}{time} was  2.9 hr using the   G5304 (B600) grating.  The slit  length \textcolor{black}{was} 330 arcsec $\times$ 1.5  arcsec with the central wavelength set to  550.0 nm.  Further details of the instrumental set up are given in Table \ref{table2}. \textcolor{black}{Data reduction was carried out using the Gemini software package version 1.13 within IRAF. The reduction  included bias subtraction, flat-field correction, wavelength calibration and sky subtraction. Observations of Fergie 34 were used to make the  flux calibration and the spectrum's wavelength was calibrated from  a CU--Ar   lamp spectrum.} The spectrum covers  a rest frame  wavelength range \textcolor{black}{from} $\sim$ 4000 \angs\ to 6900 \angs\  with  a  resolution of R$_{B600}$ = 1688. The slit was placed along the  TDG candidate's major axis, (PA $\sim$ 59.1\degree), inferred from NUV \textit{(GALEX)} imaging and centred at the position of the TDG's NUV intensity maximum, see Figure \ref{fig3}(d). \textcolor{black}{We did not use the observation obtained on January 8th 2016, because  [OIII]$\lambda$5007 emission line fell within a  gap between the GMOS detector chips  precluding an accurate flux determination.  }

\begin{table}
\centering
\begin{minipage}{110mm}
\caption{Arp\,202 TDG GMOS observation parameters}
\label{table2}
\begin{tabular}{lll}
\hline
Property & Unit  & value         \\ 
\hline
\hline

Slit centre& RA [h:m:s]& 09:00:09.300\\
Slit centre& DEC [d:m:s]&+35:43:38.000 \\
Slit length&[arcsec]&330 \\
Slit width&[arcsec]&1.5\\
PA &[\degree]&59.1\\
Grating Optics&--&B600 G5304 array\\
Central Wavelength (rest frame)&--&550.0 nm\\
Filter &--&gg455\\
Spectral Resolution&R&1688\\
Spatial Binning&--& 2\\
Spectral Binning&--&2\\
Pixel Size in Spatial Direction&arcsec& \textcolor{black}{0.1454} \\
Pixel Size in Spectral Direction&nm &0.093\\
\hline
\end{tabular}
\end{minipage}
\end{table}

\section{Observational Results} 
\label{results}
\subsection{\han\ imaging (HCT)}
\label{result_ha}
\begin{figure*}
  \begin{center}
  
  \includegraphics[angle=0,scale=0.69]{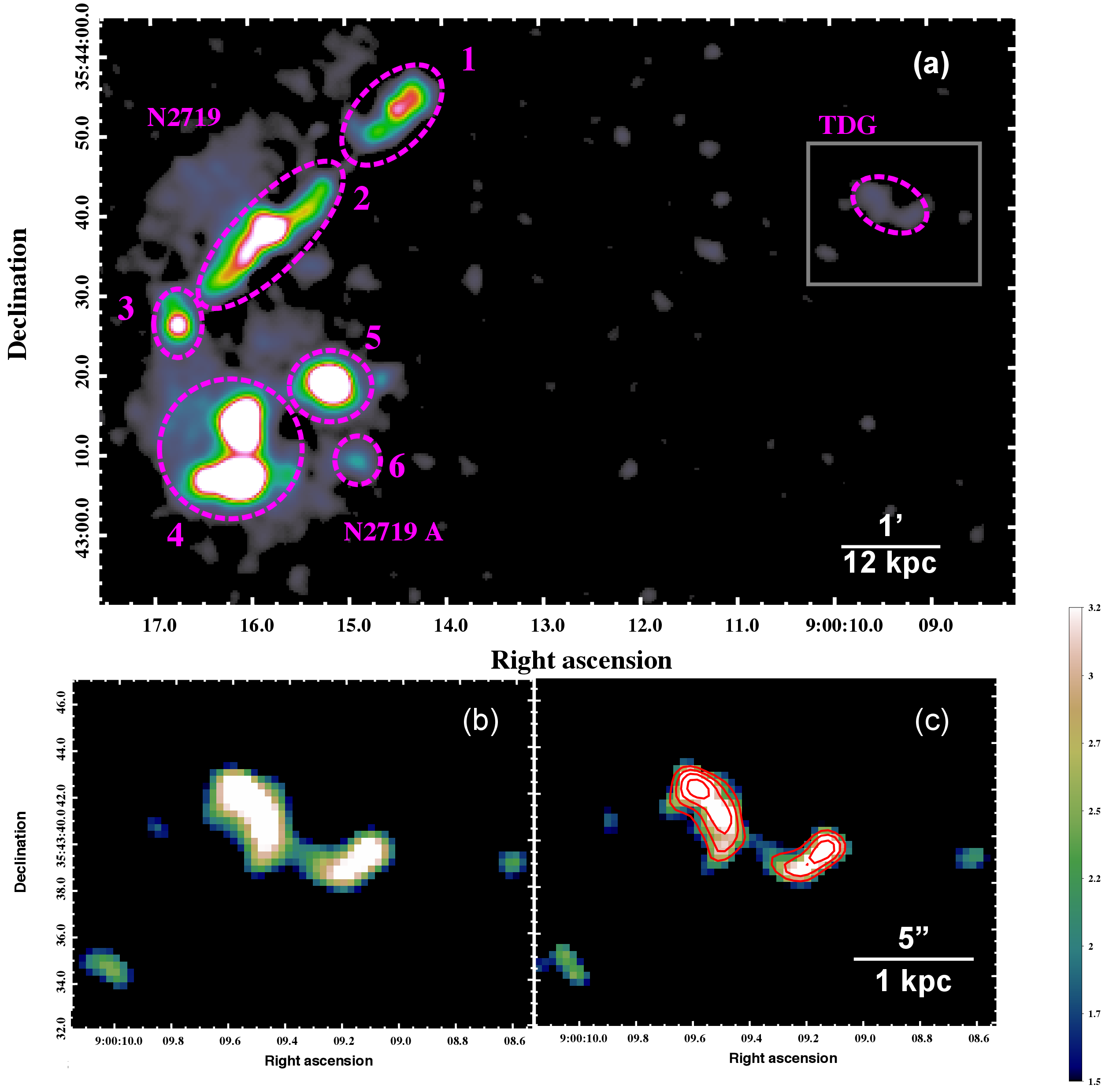}
  \caption{\textcolor{black}{\textbf{Continuum subtracted \han\  images of the Arp\,202 system. \textit{Panel (a):}} identifies 6 star forming regions  in the parent galaxies and the TDG, projected  83 arcsec ($\sim$ 17 kpc) to the  south--west of the parent pair. Star forming knots numbered 1, 2 and 3 are associated with galaxy NGC\,2719 and knots 4, 5 and 6 are associated with NGC\,2719A. 
\textbf{\textit{Bottom left panel (b):}} shows a zoomed in to the TDG in  \han\  image. \textit{\textbf{Bottom right panel (c):}} shows the signal--to--noise map for the  TDG with the contours drawn at 2, 2.5, 3.0, 3.5 $\sigma$ above the background and the signal--to--noise colour scale is shown on the right. 
}}
  \label{fig2}
  \end{center}
\end{figure*}

The  \textcolor{black}{major SF} zones traced  by \han\ emission in Figure \ref{fig2}(a) are not well isolated, with their extensions making it difficult to define the border of each  individual zone. It is likely that the undetected \han\ emission from less luminous  H{\sc ii} clumps, in the vicinity of the detected \han\ emitting regions, extend beyond their  apparent boarders \citep{rozas1999}. Hence,  we group  several H{\sc ii} zones into one star forming complex, which we refer to from here on as    ``SF regions". Following \cite{ramya2007}, we selected the \han\ emission from SF regions based on the criteria that the emission is centrally peaked; the boundary of the region is set where the flux falls  to  2\sig\ compared to the background (i.e. above $\sim10^{-17}$ \ergs). 
\textcolor{black}{The dashed ellipses in Figure \ref{fig2}(a)  enclose the boundary where the \han\ emission falls below the 2$\sigma$ background for  six star forming regions surrounding NGC\,2719 and NGC\,2719A, as well as   a counterpart to the TDG candidate, identified by \citep{smith10a}, \textcolor{black}{projected} $\sim$ 83 arcsec (17 kpc) NW of NGC\,2719A. The total \halpha\ fluxes for the TDG and the other star forming regions are estimated from  the \han\ flux within the ellipses.} The  SF regions, numbered 1, 2 and 3, in the image are projected against \textcolor{black}{or near}  the edge--on optical disk of NGC\,2719, while knots 4, 5 and 6 are projected against \textcolor{black}{or near}  the optical disk of NGC\,2719A. An area of low density \han\ emission  between knots 3 and 4 may be part of a tidal bridge between the principal pair. Figure \ref{fig2}(a)  also shows an extensive area of low surface brightness emission  N of the NGC\,2719 optical centre. 

\begin{figure*}
\begin{center}
\includegraphics[angle=0,scale=0.6]{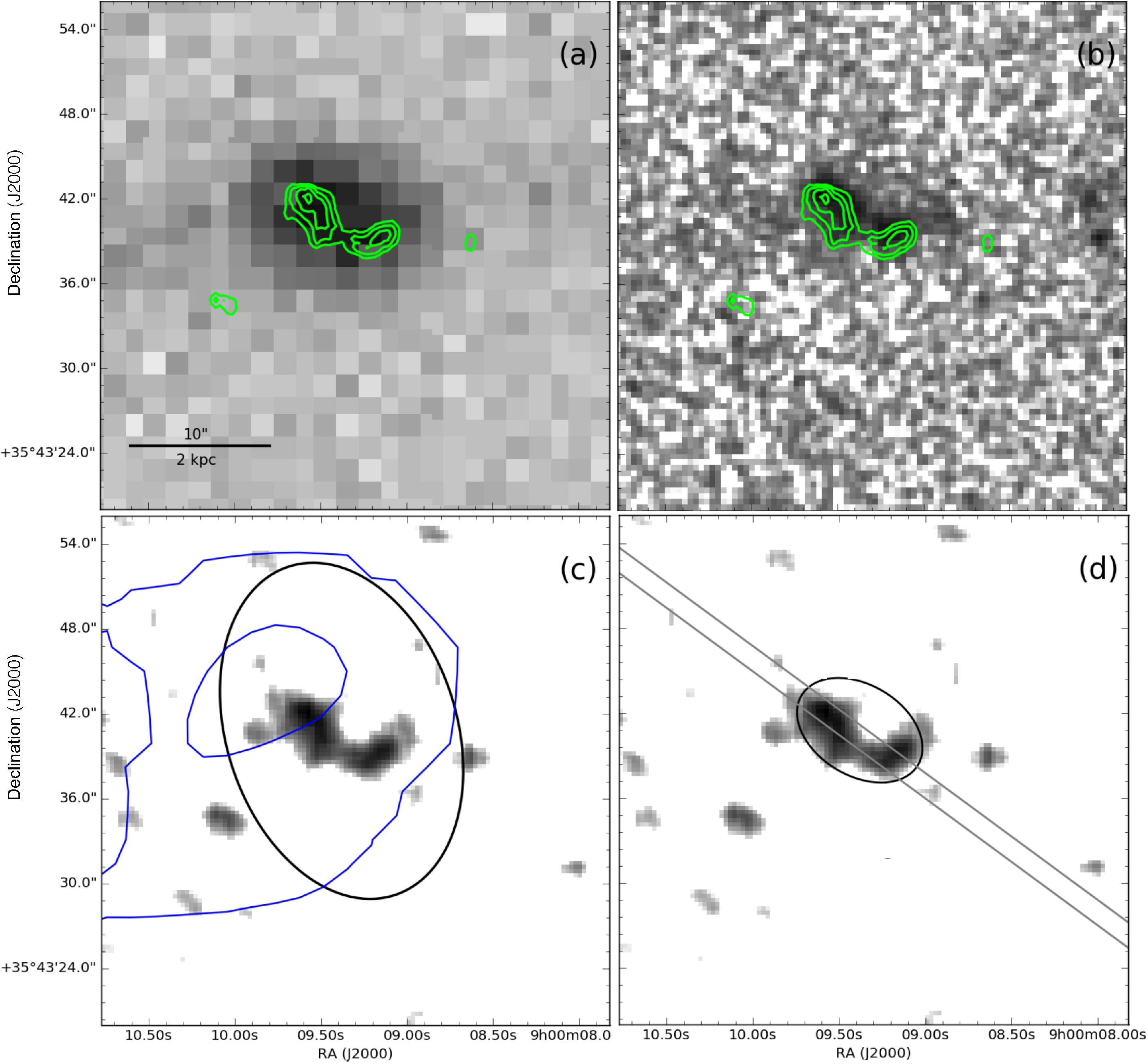}
\caption{\textbf{Arp\,202 TDG:} \textit{\textbf{(a): } } Contours (green) from the  smoothed   \han\ (HCT) image on \textcolor{black}{a} NUV(\textit{GALEX}) \textcolor{black}{gray scale} image. The \textcolor{black}{contours are at 3, 4, 5 and 6} $\sigma$ above the background level in the smoothed image. \textit{\textbf{(b): } } Contours (green) from smoothed  \han\ image on an SDSS g -- band image. \textit{\textbf{(c): } } \hi\ (GMRT)  contours overlaid on a smoothed  \han\ image. Contours are at the same levels as in Figure \ref{fig1}, with the ellipse showing the orientation and size of the GMRT FWHP (23.4 arcsec $\times$ 16.3 arcsec) synthesised beam. \textit{\textbf{(d):} } Smoothed \han\ image. The ellipse shows the  9.6" $\times $ 6.4" (1.9 kpc $\times $ 1.3 kpc) region from which the \han\ flux in table \ref{table3} was integrated  with the parallel lines indicating GMOS slit width and  orientation. The \han\ images and contours are from an image   smoothed with a boxcar (6 $\times$ 6) filter. }
\label{fig3} 
\end{center}
\end{figure*}

\textcolor{black}{ We detect a faint system,  above the 2$\sigma$ background, in the \han\ image, Figures \ref{fig2}(b) and (c), at the position of the TDG candidate. Figure \ref{fig2}(b) shows the flux distribution of TDG galaxy and Figure \ref{fig2}(c) shows the signal--to--noise (S/N) map for the TDG with  contours at  2, 2.5, 3 and 3.5 $\sigma$  above the un--smoothed background.    The TDG's \han\ structure detected within the ellipse in Figure \ref{fig2}(a) has a major axis extent of $\sim$ 9.6 arcsec (1.92 kpc). Both Figure \ref{fig2}(b and c) and the green contours from a  smoothed version of \han\ image in Figures \ref{fig3} (a, b)  show an elongated structure with two principal maxima separated by $\sim$ 5.5 arcsec (1.1 kpc). Within this structure a ridge of emission extends $\sim$ 3 arcsec (0.6 kpc) S from the northern maxima. There is a small isolated \han\ region in Figure \ref{fig3}(b) with a NUV counterpart to the W of the main \han\ structure, which hints at a more extended and low surface brightness \han\ emission beyond the main detected TDG structure}

Integrated \han\ flux, \textcolor{black}{\halpha\ luminosity, SFR(\halpha)} and dimensions  for each  SF region and  the  TDG are presented in Table \ref{table3}.  \textcolor{black}{\halpha\ luminosities and \textcolor{black}{SFR(\halpha)s} in the table are calculated after correcting the observed \han\ fluxes  for  [NII] contamination. For  both SF region 5 and the TDG the [NII] fraction of the observed \han\ flux  was  0.04, based on their \textcolor{black}{respective} SDSS and GMOS spectra. We also  applied  this same [NII] fraction \textcolor{black}{correction} when calculating the \halpha\ luminosity and SFR(\halpha)  for the remaining SF regions, which do not have available spectra. An extinction correction (logarithmic reddening parameter c(H$\beta$) =  0.04) was applied in calculating L(\halpha) and SFR(\halpha) for SF region 5  based on measurements from its SDSS spectrum. No extinction correction was made for the TDG because the c(H$\beta$)  derived from its   GMOS spectrum\textcolor{black}{, using \halpha/H$\beta$,}  was   0.00$\pm$0.01. We did not make any  extinction correction for the other SF regions.} The table's SFR(\halpha) values   are based on the formula below derived by  \cite{kennicutt1998} for galaxies with solar abundances and an assumed  Salpeter IMF with stellar masses ranging from 0.1 -- 100 \msolar:
\begin{equation}
 \footnotesize{{\rm SFR(H\alpha)} \, ({\rm M}_\odot\,{\rm yr}^{-1}) = 7.9\times10^{-42}\,L(\rm{H}{\alpha}) \, (erg\,s^{-1})}
\end{equation}


The integrated flux, F(\han), for the TDG from the image is 1.06$\pm$0.98$\times\ $10$^{-15}$ \ergs,  corresponding to an \textcolor{black}{L(\halpha) = 2.08$\pm$1.92 $\times\ $10$^{38}$ erg\,s$^{-1}$ and SFR(\halpha) =  $0.002\pm0.002$ \msolaryr. }

For the individual SF regions  in  NGC\,2719 \textcolor{black}{L(\halpha) is} in the range $\sim$ 21 -- 93 $\times10^{38}$ erg\,s$^{-1}$ (refer Table \ref{table3}). Their estimated SFR(\halpha)s are in the range \textcolor{black}{0.02 -- 0.07} \msolaryr. Similarly, for NGC\,2719A  the estimated individual SF region \textcolor{black}{L(\halpha)} and  SFR(\halpha)s  are in the range \textcolor{black}{6} -- 295 $\times10^{38}$ erg\,s$^{-1}$ and 0.004 -- 0.23 \msolaryr, respectively.

\begin{table*}%
\begin{minipage}{150mm}
\begin{center}
\caption{\textcolor{black}{\textbf{Arp 202 SF regions:} properties derived from HCT imaging. }}
\label{table3}
\begin{tabular}{lrrrrr}
\hline
SF region. & F(\han) & \textcolor{black}{L(\halpha)\footnote{\textcolor{black}{No extinction correction has been made, except  for SF region 5 where the correction is based on c(H$\beta$) =  0.04.   }}} & SFR(\halpha)\footnote{SFR(\halpha)  per equation 1 with L(\halpha) = L(\han) $\times$ 0.96, where 0.96 is \halpha\ fraction determined from both the TDG GMOS spectrum for the TDG and SF region 5 from its SDSS spectrum.}  & \textcolor{black}{Extent\footnote{\textcolor{black}{The major and minor axis of the ellipse drawn around each star forming Region.}}} \\
	     &  $10^{-15}$erg\,cm$^{-2}$\,s$^{-1}$ & $10^{38}$erg\,s$^{-1}$  & \msolaryr  & kpc $\times$ kpc \\
\hline
\hline
TDG	&	1.06$\pm$\,\,\,0.98	&\,	2.08$\pm$\, 1.92	&	0.002$\pm$0.002	&	1.92 $\times $ 1.30\\
\hline								
1	&	16.46$\pm$\,\,\,3.87	&	32.23$\pm$\, 7.57	&	0.025$\pm$0.006	&	2.99 $\times $ 1.56\\
2	&	47.97$\pm$\,\,\,6.60	&	93.95$\pm$12.92	&	0.074$\pm$0.010	&	4.83 $\times $ 1.76\\
3	&	11.06$\pm$\,\,\,3.17	&	21.65$\pm$\, 6.20	&	0.017$\pm$0.005	&	1.61 $\times $ 1.04\\
4	&	150.58$\pm$11.69	&	294.89$\pm$22.89	&	0.233$\pm$0.018	&	3.20 $\times $ 3.08\\
5	&	49.81$\pm$\,\,\,6.73	&	112.8$\pm$15.57	&	0.089$\pm$0.012	&	1.93 $\times $ 1.76\\
6	&	2.86$\pm$\,\,\,1.61	&	5.61$\pm$\, 3.16	&	0.004$\pm$0.002	&	1.31 $\times $ 1.30\\

\hline
    
\end{tabular}
\end{center}

 \end{minipage}{}
\end{table*}




\subsection{TDG spectrum (GMOS)}
\label{result_gmos}

\begin{figure*}
\begin{center}
\includegraphics[scale=0.9]{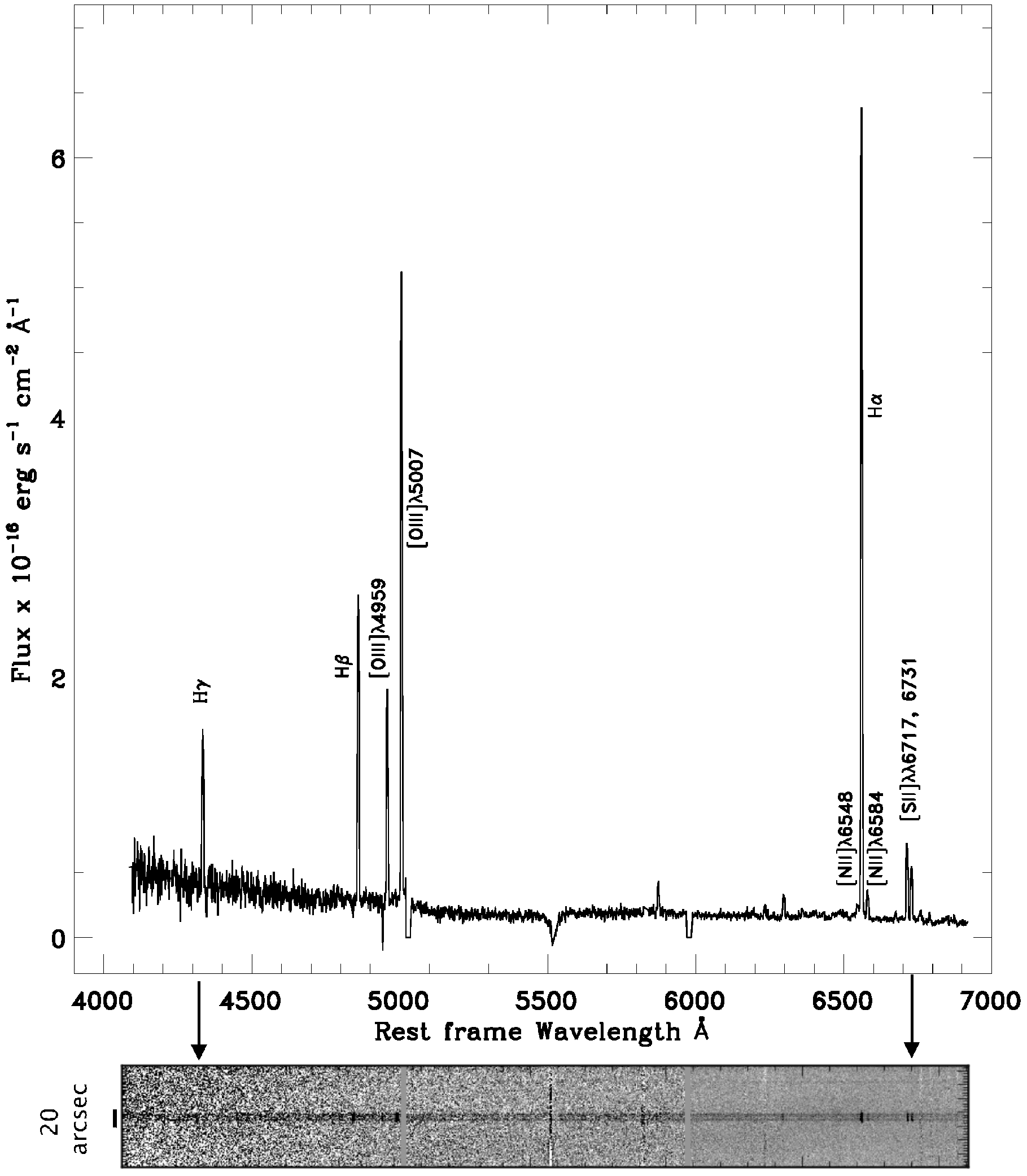}
\caption{\textbf{Arp\,202 TDG GMOS spectrum:} \textcolor{black}{\textbf{\textit{Above: }}1D spatially integrated spectrum, with the principal emission lines labelled. \textbf{\textit{Below:}} The 2D spectrum with single headed arrows identifying the  H$\gamma$ and [SII] doublet emission lines. Only the inner part of the slit spatial axis is shown, with the solid bar to the left of the 2D spectrum indicating the spatial extent (20 arcsec) resolved in the line profiles in Figure \ref{fig5}. }}
\label{fig4_0} 
\end{center}
\end{figure*}

\textcolor{black}{Figure \ref{fig4_0} shows the \textcolor{black}{GMOS 1D and 2D spectra }  of the TDG and identifies the main emission lines detected and used in our study. Emission line profiles across the slit and } integrated emission line fluxes  for  the TDG from the GMOS spectrum were measured by fitting Gaussian profiles with the IRAF task {\sc fitprofs}. These emission line fluxes  were corrected for extinction using the observed Balmer decrement. Then, the logarithmic reddening parameter c(H$\beta$)  was calculated from the ratio of H$\alpha$/H$\beta$ \textcolor{black}{and H$\gamma$/H$\beta$, where  intrinsic values of 2.86 and 0.47 for case B at 10$^{4}$ K recombination were} assumed \citep{oster06}. Therefore, the corrected emission line fluxes were calculated as:
\begin{equation}
 I(\lambda)/I(H\beta)\, =\, F(\lambda)/F(H\beta)\,  \times 10^{c(H\beta)f(\lambda)}
 \end{equation}
where I($\lambda$) and F($\lambda$) are the de--reddened flux and observed flux at a given wavelength, respectively, and f($\lambda$) is the reddening function given by \cite{cardelli89}. \textcolor{black}{The total TDG emission line fluxes relative to H$\beta$ from the GMOS spectrum are set out in Table \ref{table4}, but because  c(H$\beta$) from H$\alpha$/H$\beta$ is 0.00 $\pm$0.01,  we assume the extinction obtained from  H$\gamma$/H$\beta$. }  

\begin{table}%

\begin{minipage}{75mm}

\begin{center}
\caption{Arp\,202 TDG: integrated GMOS emission line fluxes  relative
 to H$\beta$  }
\label{table4}
\begin{tabular}{lrrr}
\hline
 Emission line &  F($\lambda$)/F(H$\beta$) \textcolor{black}{$\times$ 100} &  \\
 \textcolor{black}{(rest frame)}\\
 \hline
 \textcolor{black}{H$\gamma$\,\,\,\,\,\,\,\,\,\,\,$ \lambda$4340}&\textcolor{black}{49.63$\pm$4.03}\\
H$\beta$\,\,\,\,\,\,\,\,\,\,\,$\lambda$4861              &100.00$\pm$\textcolor{black}{4.53} \\
$\left[OIII\right]\, \lambda$4959   & \textcolor{black}{67.26}$\pm$\textcolor{black}{3.44} \\
$\left[OIII\right]\, \lambda$5007   &\textcolor{black}{196.71}$\pm$\textcolor{black}{7.04}\\
H$\alpha\,\,\,\,\,\,\,\,\,\, \lambda$6563             &\textcolor{black}{257.42}$\pm$\textcolor{black}{4.44} \\
$\left[NII\right]\,\,\,\lambda$6584    & \textcolor{black}{10.43}$\pm$\textcolor{black}{0.88} \\
$\left[SII\right]\,\,\,\, \lambda$6717    & \textcolor{black}{42.46}$\pm$\textcolor{black}{1.71} \\
$\left[SII\right]\,\,\,\, \lambda$6731    & \textcolor{black}{19.89}$\pm$\textcolor{black}{1.20} \\
\hline
    F(H$\beta$) erg cm$^{-2}$ s$^{-1}$& \textcolor{black}{1.77}$\pm$\textcolor{black}{0.03} $\times$ 10$^{-15}$ \\
\hline
c(H$\beta$)   \textcolor{black}{using H$\alpha$/H$\beta$ }                                                               &    \textcolor{black}{0.00$\pm$0.01}     \\ 
c(H$\beta$)   \textcolor{black}{using H$\gamma$/H$\beta$}                                                               &    \textcolor{black}{0.03$\pm$0.01}     \\

\hline
\end{tabular}
\end{center}
\end{minipage}{}

\end{table}

From the spectrum we derive  L(\halpha) = \textcolor{black}{9.45} $\times$ 10$^{38}$ erg s$^{-1}$, EW(\halpha) = 2052 \angs\ and EW(H$\beta$) = \textcolor{black}{89} \angs\ for the TDG. The [SII]  6717$\lambda$/6731$\lambda$ flux ratio for the TDG from the fluxes in  Table \ref{table4} is \textcolor{black}{2.13}, assuming T$\sim$10$^4$ K and implies   n$_e \sim$ 100 cm$^{-3}$. We estimate \textcolor{black}{the mass of the ionized gas}  using the equation:\\
\begin{equation}
M(HII) =\frac{L(H\alpha)m_{p}}{n_{e}\alpha_{H\alpha}^{eff}h\nu_{H\alpha}}
\end{equation}

\noindent where $m_p$ = the proton mass, $n_{e}$ = the electron density and $\alpha_{H\alpha}$ = the effective recombination coefficient, respectively. \textcolor{black}{Applying this equation  we derive M$_{HII}$ = \textcolor{black}{2.20$\pm$0.07 $\times$ 10$^4$M$_{\odot}$} and conclude that ionized}  gas makes an insignificant contribution to the total mass of the TDG\textcolor{black}{, see Table \ref{table_1}}.

\begin{figure*}
\begin{center}
\includegraphics[scale=0.63]{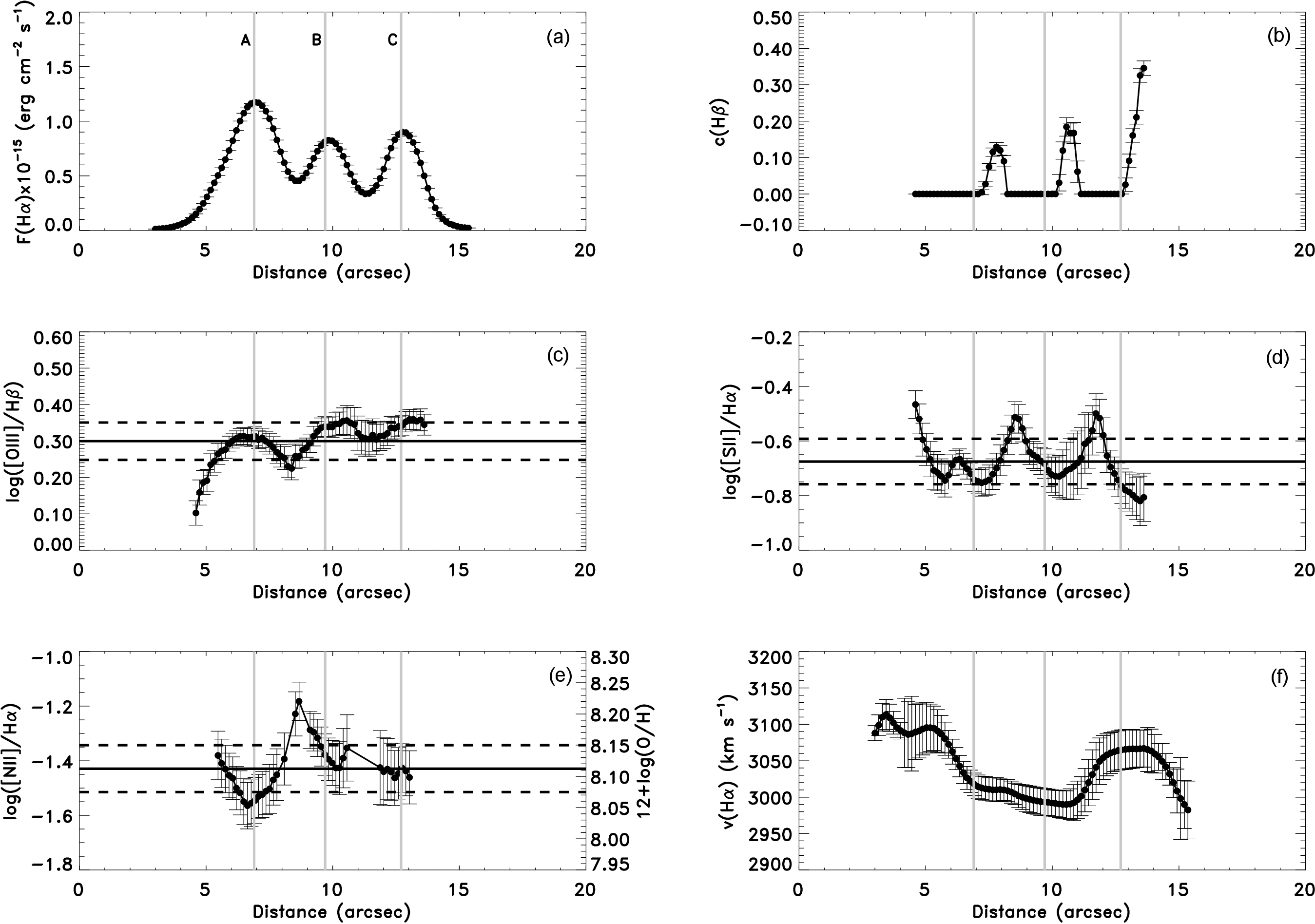}

\caption{\textbf{\textbf{Arp\,202 TDG GMOS spectrum profiles \textcolor{black}{across} the slit}:} \textbf{\textit{(a):}} \halpha; \textbf{\textit{(b):}} c(H$\beta$);  \textbf{\textit{(c):}} log([OIII]/H$\beta$); \textbf{\textit{(d):}} log([SII]/H$\alpha$); \textbf{\textit{(e):}} log([NII]/H$\alpha$), \textbf{(f):} v(H$\alpha$) with the TDG's systemic H$\alpha$ velocity = \textcolor{black}{3032} \km. The  maxima of three principal \halpha\ knots (A, B, C) \textcolor{black}{across}  the slit are marked with vertical gray lines. \textcolor{black}{Solid horizontal lines over plotted on the profiles indicate the mean values with the 1 $\sigma$ values indicated with dashed lined }. The error bars are the 1 $\sigma$ uncertainties. 1 \textcolor{black}{arcsec} $\sim$  200 pc. }
\label{fig5} 
\end{center}
\end{figure*}

Figure \ref{fig5} shows the Arp\,202 TDG GMOS spectrum profiles \textcolor{black}{across} the slit for: \textbf{\textit{(a):}} \halpha;  \textbf{\textit{(b):}} c(H$\beta$);  \textbf{\textit{(c):}} log([OIII]/H$\beta$); \textbf{\textit{(d):}} log([SII]/H$\alpha$); \textbf{\textit{(e):}} log([NII]/H$\alpha$) \textcolor{black}{and } \textbf{\textit{(f):}} v(H$\alpha$). \textcolor{black}{Because the H$\beta$ line is more extended than H$\gamma$, which is only resolved in a few pixels, we show the \halpha/H$\beta$ extinction profile in   Figure \ref{fig5}(b).    } The three principal knots \textcolor{black}{across} the slit are marked \textcolor{black}{with vertical gray lines (A, B and C) } on each profile.

Figure \ref{fig5}(b) shows the  c(H$\beta$) profile across the TDG. The reddening parameter c(H$\beta$) was set to 0.0 for unrealistic   H$\alpha$/H$\beta$ values $<$ 2.86. We inferred a range of extinction from $\sim$0.0 to \textcolor{black}{0.35}  in the spectrum. It is interesting to note from Figures \ref{fig5}(a and b) that the local   H$\alpha$ emission and extinction maxima do not coincide spatially. This suggests that the current or a recent star--burst episodes  are sweeping the gas and dust out of the TDG \textcolor{black}{centre} into the surrounding regions. This same phenomena has been  also been observed in some \textcolor{black}{other star--forming} dwarf galaxies with single and multiple SF regions \citep[e.g.][]{lagos09,lagos12,lagos14,lagos16}. 

\begin{figure}
\begin{center}
\includegraphics[scale=0.6]{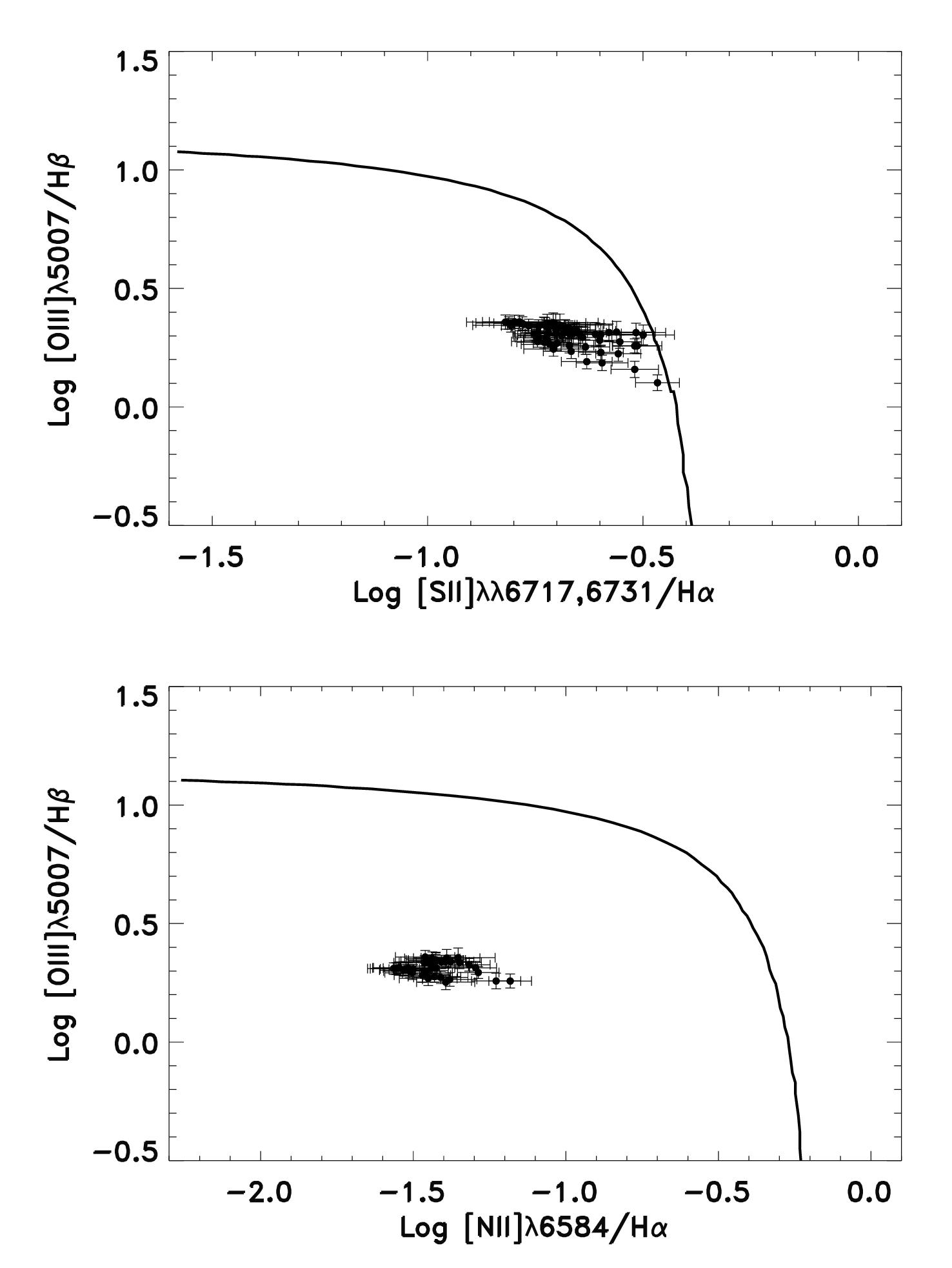}
\caption{\textcolor{black}{\textbf{Arp\,202 TDG BPT diagrams.} \textbf{\textit{Top:}} [O III]$\lambda$5007/H$\beta$, [S II]$\lambda\lambda$6717,6731/H$\alpha$. \textbf{\textit{Bottom:}} [O III]$\lambda$5007/H$\beta$, [N II]$\lambda$6584/H$\alpha$. The area to the left of the solid line is the area in the diagrams predicted for young stars by photo--ionization models.} }
\label{fig4a} 
\end{center}
\end{figure}

To infer the dominant ionization source in the TDG we employed the commonly used diagnostic or BPT \citep{BPT1981} diagrams \textcolor{black}{(Figure \ref{fig4a})} using the following emission-line ratios:  [O III]$\lambda$5007/H$\beta$, [S II]$\lambda\lambda$6717,6731/H$\alpha$ and [N II]$\lambda$6584/H$\alpha$. In Figures \ref{fig5}(c, d, e) we show those emission--line--ratio profiles.
\textcolor{black}{The mean value of each emission line ratio is log([OIII]/H$\beta$) = 0.30, log([SII]/\halpha) = -0.67 and log([NII]/\halpha) = -1.43, with a standard deviations of  0.05, 0.08 and 0.09, respectively. The profiles of these emission line ratios changes from the peak of \halpha\ emission to the outer part of each SF knot. The log([OIII]/H$\beta$) ratio decreases, while log([SII]/\halpha) and log([NII]/\halpha) ratios  increases with distance \textcolor{black}{from the \halpha\ maxima}. This figure also shows that the ionization structure of the SF \textcolor{black}{knots}, in the TDG, is rather regular at 1$\sigma$ level for all the ionization emission line ratios. However, the \textcolor{black}{maximum} in the  log([NII]/\halpha) profile is more than 2$\sigma$ above the mean value. All points in the BPT diagram (Figure \ref{fig4a}) fall within the locus predicted for young stars by photo-ionization models in the BPT diagrams \citep{oster06}.}  Therefore, we conclude  photoionization from stellar sources is the dominant excitation mechanism in the TDG. The regular emission line profiles, which can be fitted with a single  Gaussian component, argue against any significant \textcolor{black}{fast} shocks within the GMOS slit. \oab\ abundance, Figure \ref{fig5}(e), is derived by applying the relation between the line ratio of [N II]$\lambda$6584/H$\alpha$ with the oxygen abundance from \cite{dene02}, i.e., \oab\ = 9.12 + 0.73 $\times$ N2, with N2 = log([N II] $\lambda$6584/H$\alpha$). This gives an  integrated value of \oab\ = 8.10\textcolor{black}{$\pm$0.41} for the TDG. Applying the same procedure to  the  SDSS spectrum for SF region 5  in NGC\,2719A  we obtain a value of \oab\ = 8.05\textcolor{black}{$\pm$0.41}. Figure \ref{fig5}(f) shows the \halpha\ velocity, v(\halpha), profile \textcolor{black}{across}  the slit. The TDG's mean  v(\halpha) = \textcolor{black}{3032} \km, confirming its velocity is close to those of that of the  parent galaxies as reported by \cite{smith10a}. 

Using Equation 1 and the \halpha\ emission determined from the GMOS spectrum we estimated the total SFR for the TDG as   SFR(\halpha) = \textcolor{black}{0.007$\pm$0.001} M$_{\odot}$ yr$^{-1}$, which is in  \textcolor{black}{reasonable}  agreement with the SFR(\halpha)  derived from HCT imaging of 0.002 \msolaryr, see section \ref{result_ha}. Ordinarily we would expect the TDG \halpha\ flux derived from narrow band imaging to exceed the value from a long slit spectrum because it integrates flux from a more extensive area. But as noted in section \ref{obs-hct} the sky conditions for during the \han\ imaging only allowed the  lower limit for the TDGs L(\han) and SFR(\halpha) to be determined from the HCT observations.

\section{Discussion}
\label{dis}
In the following section we discuss the relationship between the parent pair's interaction and the formation of the TDG.   We were particularly interested to see whether the \textcolor{black}{morphological, chemical and kinematic  evidence supports} the candidate being a true TDG and, if so, whether its star formation history (SFH) is consistent with its stars being formed in--situ under the pure gas collapse scenario  as the Arp\,202 \hi\ and UV  morphologies suggest.

\subsection{NGC 2719 and NGC 2719A}
\label{dis2719A}
The optical radial velocity separation of the Arp\,202 pair, NGC 2719 and NGC\,2719A, is only 40 \km. Respectively, their optical  D$_{25}$ is 1.3 arcmin (15.6 kpc) and 0.9 arcmin (11 kpc) with respective M$_*$ \textcolor{black}{values} of  $\sim$3.4 $\times$ 10$^9$ \msolar\ and 5.0 $\times$ 10$^9$ \msolar\  \citep{seng14}. These properties indicate both members of the pair are  small lower mass \textcolor{black}{galaxies} with a M$_*$ ratio of $\sim$ 1:1.5. The  \han\   bridge seen in Figure \ref{fig1} between the eastern ends of the pair is also seen at  lower resolution in the 3.6 $\mu$m, \hi\ and  NUV images  in the same figure.  \textcolor{black}{This near--side tidal bridge between the galaxies   and the  far--side}  tidal tail \textcolor{black}{ seen in Figures \ref{fig1}b and \textcolor{black}{\ref{fig1}}c  are typical morphologies for tidally interacting pairs}  \citep{struck99,oh08}. The \textcolor{black}{pair's}  highly perturbed \hi\ morphology and kinematics  \citep{seng14} indicate the  parents' \textcolor{black}{most recent} periapse occurred  well within the 0.4 Gyr  to 0.7  Gyr during which \hi\ interaction signatures are predicted to remain detectable for interactions of this magnitude \citep{holwerda11}. \textcolor{black}{Analysis of an  SDSS  spectrum for a SF region within NGC\, 2917A  also provides insights into interaction between the pair.}

\textcolor{black}{The SDSS spectrum for the SF region 5 in NGC\,2719A is classed in SDSS as being from a galaxy star burst  with an EW(\halpha) = \textcolor{black}{621$\pm$14} \angs.   It is reasonable to  \textcolor{black}{assume}  this large  EW(\halpha) is the result of the most recent pair interaction, with the \halpha\ emission reflecting SF  on time scales of $\sim$10$^7$ yr. Allowing \textcolor{black}{that} there  may be a delay between the interaction \textcolor{black}{and the  onset of a}  SF burst \textcolor{black}{as well as}  recognising the \hi\ remains highly perturbed,  we concluded that the most recent} interaction between the pair probably occurred within a few  $\times$ 10$^8$ yr ago.

\subsection{TDG Candidate}
\label{disTDG}

Our analysis of  the \textcolor{black}{TDG's} GMOS spectrum gives an oxygen abundance of 12+log(O/H) = 8.10\textcolor{black}{$\pm$0.41}, which is lower  than the earlier  value of 8.9 from \cite{smith10a}. The Arp\,202 parent galaxies are lower mass \textcolor{black}{and gas rich galaxies,} so we would expect them to have somewhat sub--solar oxygen abundances and  the oxygen abundance of SF region 5 in  NGC\,2719A, derived from its SDSS spectrum (12+log(O/H) = 8.05\textcolor{black}{$\pm$0.41}),  is consistent with this expectation. The excellent agreement between the oxygen abundances in the TDG and SF region 5 in  NGC\,2719A strengthens the case for the material in the TDG having originated from the parent pair and most probably NGC\,2719A. Chemical evolution modelling by \cite{ploeck14} predicts no significant increase in a TDG's  integrated oxygen abundance from in--situ SF within  the first  5 $\times$ 10$^8$ yrs. But, their models indicate there can be a metallicity enhancement toward the centre of the TDG as is observed in the Arp\,202 TDG between knots A and B  (Figure \ref{fig5}(e)) where the  oxygen abundance reaches \oab\ $\sim$ 8.2. This value is 2$\sigma$ above the mean 12+log(O/H) abundance of $\sim$ 8.08  (log(NII/\halpha) = -1.43;  see Figure \ref{fig5}(e). \textcolor{black}{It is interesting  that opposite trend in 12+log(O/H) abundance with radius is seen in a NGC 5291N  TDG  \citep{fensch16}. However, \cite{fensch16} note that the  apparent trend \textcolor{black}{is}, within the errors,   consistent with a homogeneous metallicity distribution.  }

Corroboration of the TDG's current low  SFR, referred to in section \ref{result_gmos}, comes from  the absence of  8$\mu$m emission in the \textit{Spitzer} image presented in \cite{smith07}. The TDG's SFR(FUV)\footnote{Based on SFR(FUV) = 1.4 $\times$ 10$^{-28}$  L$_v$(UV)  and  f$_v$ = 10$^{-0.4(m_{AB}+46.8)}$ from \citep{lee09} } (0.011 M$_{\odot}$ yr$^{-1}$), derived from the FUV magnitude (m$_{AB}$ = 20.6) from \textcolor{black}{\cite{smith10a}}, is \textcolor{black}{$\sim$  50\% higher   than the SFR(\halpha) from GMOS. In general  the \textcolor{black}{derived SFR(UV) is lower than  SFR(\halpha) in star forming regions,}  with the difference attributed to dust extinction of the UV emission. But for the Arp\,202 TDG the  c(H$\beta$) is so low that extinction cannot be a  significantly contributor to the difference. Instead, the difference between  SFR(FUV) and the GMOS SFR(\halpha)   is probably due to the limited area of the TDG covered the GMOS slit and/or a rapid decline in SF within the last $\sim$ 10$^8$ yr. } \textcolor{black}{ A rapid decline in SF is compatible with the idea that high SFR's in star-forming dwarfs, such as blue compact dwarf galaxies (BCDs), cannot be sustained for long periods. For example, the depletion time scales in the BCDs Mrk 36 and UM 461 \citep{lagos11} are significantly less than a Hubble time $\sim$0.2 Gyr and 2 Gyr, respectively. These studies suggest that these objects undergo a few or several short bursts of star formation.} 

Whether or not tidal debris, including TDGs, contain  stellar populations older than the  interaction during which they formed,  provides important clues about their  formation history. For example  star forming regions within the tidal bridge between M\,81 and M\,82 contain  both young and old ($>$ 1 Gyr) stellar populations, with the old population thought to be tidal stellar debris from the interaction \citep{makarova02}. \textit{Spitzer} near infra-red  (NIR) images  at 3.6 $\mu$m and 4.5 $\mu$m trace emission from the old stellar population of a galaxy ($\gtrsim$ 1 Gyr), but they can be contaminated  by PAH and dust emission, as well as emission from  AGB and  RGB stars  formed in more recent SF bursts \citep{meidt2014}. This contamination is most severe in strongly star forming regions\textcolor{black}{, but  given the   current low  Arp 202 TDG SFR,  this contamination is expected to be minimal. Therefore, the  absence of   \textit{Spitzer} 3.6 $\mu$m and 4.5 $\mu$m counterparts,  see Figure \ref{fig1}(d) and \cite{smith07}, indicates  a low upper limit for the mass of the old stellar population. Integrating the flux in an circular aperture of 20 arcsec diameter from the TDG region we obtain total flux 8.2   and 14.1 \muJy\ with a 3 $\sigma$  sky-background 25.6 and 32.9 \muJy\ for the Spitzer 3.6 and 4.5 channels, respectively. This gives an upper limit of stellar mass of the TDG candidate $\approx$ 5 $\times$10$^{6}$ \msolar\ or  an order of magnitude lower than the stellar mass estimates from photometry of SDSS images per Table \ref{table_1}.  The absence of 3.6 $\mu$m and 4.5 $\mu$m  NIR emission, the lack of absorption features in the GMOS spectrum and the weak continuum in the GMOS spectrum all point to absence of any  significant  mass of old stars  ($>$ 1 Gyr)  in the TDG. In an effort to confirm the absence of an older stellar population we tried to apply spectral synthesis codes (STARLIGHT\footnote{\textcolor{black}{http://www.starlight.ufsc.br.}} and FADO v.1\footnote{\textcolor{black}{Fitting Analysis using Differential evolution Optimization (FADO) \citep{gomes17}.}}) to the Arp\,202 TDG GMOS spectrum. The STARLIGHT fit produced results which was inconsistent with the evidence from other wavelengths.  FADO, unlike most spectral synthesis codes, accounts  for nebular continuum and is optimised for young star forming galaxies.    While, the FADO results appeared to confirm the absence of an old stellar population (stellar ages $<$ 5 $\times$ 10$^8$ yr),  the quality of the continuum  in  the GMOS spectrum (signal to noise $\sim$ 3) is too low to produce  a credible fit. Consequently  spectral synthesis modelling of the  continuum  from the GMOS spectrum could not be used to support a claim of an exclusively young stellar population.}  

\textcolor{black}{The \halpha\ knot velocities do not appear follow a systematic trend \textcolor{black}{across the slit},  see Figure \ \ref{fig5}(f), suggesting the TDG as whole has not \textcolor{black}{yet} reached a  virialised state.}  \textcolor{black}{ While studying the effect of \hi\ on star formation, we find that the \hi\ column density, N(\hi), local maxima near  the TDG position is N(\hi) $\sim$  7.5 $\times$ 10$^{20}$ atoms cm $^{-2}$, is  above the threshold for SF, N(\hi) $>$ 4 $\times$ 10$^{20}$ atoms cm $^{-2}$ from \cite{maybhate07}.  An additional finding for the Arp\,202 TDG is that it has a  low star formation efficiency  (SFE).} \textcolor{black}{Based on the estimated \hi\ masses for the  parent galaxies (7.1$\pm$0.3 $\times$ 10$^9$ \msolar)  \citep{seng14}  and using the sum of  SFRs from Table \ref{table3}  the SFE for the Arp\,202 pair (SF regions 1 to 6) is  6 $\times$ 10$^{-11}$yr$^{-1}$.  For the TDG the estimated  SFE is 7.4 $\times$ 10$^{-11}$ yr$^{-1}$, based on its \mhi\ = 1.0$\pm$0.5 $\times$ 10$^8$ \msolar\  \citep{seng14} and SFR(\halpha) from GMOS.  The TDG's SFE  is unexpectedly  similar to that for the interacting pair. Even accepting this, the TDG's SFE is still lower than \textcolor{black}{that} calculated for the Arp\,305 Bridge TDG (3 $\times$ 10$^{-10}$ yr$^{-1}$) from \cite{seng17}.}




The location of the  \textcolor{black}{Arp\,202 TDG, \textcolor{black}{with evidence that it is} comprised of young stellar populations \textcolor{black}{with a similar metallicty to a parent},   at the extremity  of the \hi\ and UV tidal tail \textcolor{black}{(the tail itself being essentially free of  young \han\ clumps)}  is  consistent with the scenario that the TDG was formed from \hi\ stripped from a parent \textcolor{black}{and within the parent's}  extended DM halo  as proposed by \citep{bournaud03,duc04}. This  scenario predicts a TDG with stellar population  consisting  entirely of stars created in--situ following  gravitational collapse of the \hi\ accumulated at the tidal tail tip, i.e., the stars will be younger than the parent pair's  interaction which produced the tidal tail and  TDG.  \textcolor{black}{Evidence supporting the absence of an old stellar population in the Arp 202 TDG comes from  \textit{Spitzer } 3.6 $\mu$m and 4.5 $\mu$m non--detection, absence of absorption features  in the GMOS spectrum, and weak continuum in the GMOS spectrum.  } } An alternative to a purely gravitational collapse of the tail tip gas is that compressive tides promoted  gas concentration prior to its collapse  \citep{renaud09,renaud15}. Compressive tides could  be an explanation for the elongated structure of the Arp\,202 TDG and the projected angular offset ($\sim$40\degree) between the TDG's  major axis  and the  tidal tail. \textcolor{black}{However} it would require pair specific  modelling to determine whether compressive tides  are likely to have played a role in the TDG's formation.  Based on the \textcolor{black}{observational} evidence to date, we conclude that a scenario in which the Arp\,202 TDG  formed from collapse of a large mass of  gas accumulated at the tip of a tidal tail following the interaction between the parent galaxies is viable.

\section{Summary and concluding remarks}
\label{summary}
We have imaged  the Arp\,202 system, including its TDG candidate, in \han\ and analysed \textcolor{black}{the TDG's}   GMOS spectrum. The GMOS observations reveal \textcolor{black}{the TDG has} recessional V$_{H\alpha}$ = \textcolor{black}{3032} \km, close  the parent pair's velocities and  an oxygen abundance (12+log(O/H) = 8.10\textcolor{black}{$\pm$0.41}). The TDG's oxygen abundance is \textcolor{black}{in good agreement with that from SF region 5 in NGC\,2719A, one of the parent galaxies, where 12+log(O/H) = 8.05\textcolor{black}{$\pm$0.41}.  Both of these TDG properties} provide positive results for key TDG validation tests  \textcolor{black}{and confirm the previous redshift measurement by \cite{smith10a}. However the derived oxygen abundance for the TDG  was lower than the 12+log(O/H) = 8.9 reported in \cite{smith10a}.} 
The \han\ imaging reveals the TDG to have an elongated  structure  $\sim$ 1.92 kpc in length with \textcolor{black}{its} two principal knots at either end of the structure. 

\textcolor{black}{The location of the TDG at the extremity  of the \hi\ and UV tidal tail and the evidence that it lacks \textcolor{black}{a significant} old stellar population  \textcolor{black}{is consistent with the Arp\,202 TDG having been formed, within the  extended DM halo of one of its patent galaxies}, under the pure gas collapse  senario proposed by \cite{bournaud03} and \cite{duc04}. \textcolor{black}{While the evidence to date does not prove this scenario, it is consistent with it. We emphasise the DM halo referred to here is that of one of the parent galaxies rather than the DM halo of the TDG itself.} However,  spectroscopic studies of a sample of TDGs are required to understand the extent to which the properties \textcolor{black}{of} TDGs formed in this way differ  from cases where the TDGs contain  a significant mass of old} stellar tidal debris. The SFHs, chemical abundances and kinematics derived  from  long slit spectroscopy would assist in  answering  this question. However the spatial distribution of properties within  TDGs  is also an important aspect of the question, requiring integral field spectroscopy (IFS). Such studies have the potential to provide insights into the impacts  of initial gravitational potential and  chemical abundance  on the subsequent SFHs and the chemical evolution of dwarf galaxies, including TDGs.    

\section{Acknowledgments}
\textcolor{black}{We are grateful to the anonymous referee for their helpful comments that have  improved the paper.} This work was supported by Funda\c{c}\~{a}o para a Ci\^{e}ncia e a Tecnologia (FCT) through national funds (UID/FIS/04434/2013) and by FEDER through COMPETE2020 (POCI-01-0145-FEDER-007672). TS acknowledges the support by the fellowship SFRH/BPD/103385/2014 funded by FCT (Portugal) and POPH/FSE (EC). P.L. is supported by a Post-Doctoral grant SFRH/BPD/72308/2010, funded by FCT (Portugal). J.--H.W. acknowledges the support by the National Research Foundation of Korea (NRF) grant funded by the Korea government (\textcolor{black}{No. 2016R1A2B3011457 and No.2017R1A5A1070354}). \textcolor{black}{Paudel S. acknowledges the support by Samsung Science \& Technology Foundation under Project Number SSTF-BA1501-0. RS kindly acknowledges the support of NSFC grant (Grant No. 11450110401) and President’s International Fellowship Initiative (PIFI) awarded by the Chinese Academy of Sciences. We also wish to thank Jean Michel Gomes and Leanadro Cardoso for assistance with spectral synthesis codes.    }  We thank the staff of the {\it HCT} and Gemini North who have made these observations possible. This research has made use of the NASA/IPAC Extragalactic Database (NED) which is operated by the Jet Propulsion Laboratory, 
California Institute of Technology, under contract with the National Aeronautics and Space Administration. 
This research has made use of the Sloan Digital Sky Survey (SDSS). Funding for the SDSS and SDSS--II has been provided by the Alfred P. Sloan Foundation, the Participating Institutions, the National Science Foundation, the U.S. Department of Energy, the National Aeronautics and Space Administration, the Japanese Monbukagakusho, the Max Planck Society, and the Higher Education Funding Council for England. The SDSS Web Site is http://www.sdss.org/.
This research made use of APLpy, an open-source plotting package for Python hosted at http://aplpy.github.com

\bibliographystyle{mnras}
\bibliography{cig}



\end{document}